\documentclass[twocolumn]{aastex6}
\usepackage{rotating,graphicx,color}

\begin{document}

%%%%%%%%%%%%%% MY DEFINITIONS
\def\xmm {\emph{XMM--Newton}}
\def\cxo {\emph{Chandra}}
\def\swift {\emph{Swift}}
\def\frm {\emph{Fermi}}
\def\igr {\emph{INTEGRAL}}
\def\sax {\emph{BeppoSAX}}
\def\xte {\emph{RXTE}}
\def\rst {\emph{ROSAT}}
\def\asca {\emph{ASCA}}
\def\hst {\emph{HST}}
\def\nst {\emph{NuSTAR}}
\def\src {\mbox{SN\,1987A}}
\def\flux {\mbox{erg cm$^{-2}$ s$^{-1}$}}
\def\lum {\mbox{erg s$^{-1}$}}
\def\nh {$N_{\rm H}$}
%%%%%%%%%%%%%%%%%%%%%%%%%%%%%%%%%%%%%%

\title{Can a bright and energetic X-ray pulsar be hiding amid the debris of \src?}

\shorttitle{A pulsar in \src}
\shortauthors{P.~Esposito et al.}

%\slugcomment{Notes on the observational constraints for a neutron star in \src}
%\date{}
%
%\watermark{Submitted}
%\setwatermarkfontsize{5cm} 
%\authorcomment2{author text}
%\paragraph{heading}
%\received{receipt date}
%\revised{revision date}
%\accepted{2018 March 12}

\author{
Paolo~Esposito,\altaffilmark{1} Nanda Rea,\altaffilmark{1,2} 
%\ldots\altaffilmark{3,4,5,6} 
Davide Lazzati,\altaffilmark{3} Mikako Matsuura,\altaffilmark{4} Rosalba Perna,\altaffilmark{5} Jos\'e A. Pons\altaffilmark{6}
}

\altaffiltext{1}{ Anton Pannekoek Institute for Astronomy, University of Amsterdam, Postbus 94249, 1090\,GE Amsterdam, The Netherlands}
\altaffiltext{2}{ Institute of Space Sciences (IEEC--CSIC), Campus UAB, Carrer de Can Magrans s/n, 08193 Barcelona, Spain}
\altaffiltext{3}{ Department of Physics, Oregon State University, 301 Weniger Hall, Corvallis, OR 97331, USA}
\altaffiltext{4}{ School of Physics and Astronomy, Cardiff University, QueenÕs Buildings, The Parade, Cardiff, CF24 3AA, UK}
\altaffiltext{5}{ Department of Physics and Astronomy, Stony Brook University, Stony Brook, NY 11794, USA}
\altaffiltext{6}{ Department de F\'{\i}sica Aplicada, Universitat d'Alacant, Ap. Correus 99, 03080 Alacant, Spain}
\email[E-mail:~]{P.Esposito@uva.nl}

\begin{abstract}
The mass of the stellar precursor of supernova (SN) 1987A and the burst of neutrinos observed at the moment of the explosion are consistent with the core-collapse formation of a neutron star. However, no compelling evidence for the presence in \src\ of a compact object of any kind has been found yet in any band of the electromagnetic spectrum, prompting questions on whether the neutron star survived and, if it did, on its properties. Starting from the analysis of recent \cxo\ observations, here we appraise the current observational situation. We derived limits on the X-ray luminosity of a compact object with a nonthermal, Crab-pulsar-like spectrum of the order of $\approx$$(1$--$5)\times10^{35}$\,\lum, corresponding to limits on the rotational energy loss of a possible X-ray pulsar in \src\ of $\approx$$(0.5$--$1.5)\times10^{38}$\,\lum. However, a much brighter X-ray source cannot be excluded if, as is likely, it is enshrouded in a cloud of absorbing matter with metallicity similar to that expected in the outer layers of a massive star towards the end of its life. We found that other limits obtained from various arguments and observations in other energy ranges either are unbinding or allow a similar maximum luminosity of the order of $\approx$$10^{35}$\,\lum. We conclude that while a pulsar alike the one in the Crab Nebula in both luminosity and spectrum is hardly compatible with the observations, there is ample space for an `ordinary' X-ray-emitting young neutron star, born with normal initial spin period, temperature and magnetic field, to be hiding inside the evolving remnant of \src.
\end{abstract}


\section{Introduction}

The supernova (SN) designated `1987A' was discovered on 1987 February 23 in the Large Magellanic Cloud (LMC). It was the brightest and nearest SN explosion observed since Kepler's SN\,1604 and is providing a wealth of information on the last evolutionary stage of massive stars as well as on the formation of a supernova remnant \citep{arnett89,mccray93,mccray16}. The explosion also confirmed the collapse of the progenitor star's core in Type II supernovae through a burst of neutrinos detected by multiple instruments \citep{hirata87,bionta87,alekseev87}.

\src\ is surrounded by a triple-ring system that formed $\sim$20 kyr before the explosion from material ejected by the progenitor, possibly as a result of its fast rotation or a binary merger \citep{morris07,chita08}. The progenitor was identified in pre-explosions images to be Sanduleak (Sk) $-$69$^\circ$\,202, which was a B3\,I blue supergiant with mass estimated  at $\sim$14\,M$_\sun$ at the time of the explosion and initially at $\sim$20\,M$_\sun$, while in the case of a binary merger, the standard model assumes two stars originally of $\sim$15 and $\sim$5\,M$_\sun$ \citep{rousseau78,gilmozzi87,hillebrandt87,sonneborn87,woosley87,walborn89,podsiadlowski92}. 

The progenitor's mass and the neutrino flash observed at the time of the SN are consistent with the birth of a neutron star, though the formation of a black hole, directly or at a later time from fallback, cannot be excluded \citep{perego15,blum16}. %Therefore, searching for a NS in \src\ was an obvious thing to do. 
So far, however, no convincing detection of a neutron star---such as  the observation of pulsed emission or of a point-like source---or compelling signs of its presence were obtained in any wavelength, and the upper limits provided by deep observations in the various bands are often perceived as ruling out the presence of a `standard' neutron star (see e.g. \citealt{mccray93,graves05,park05,manchester07,mccray07,mccray16}). 
%For example, \citet{mccray07} observed that the \cxo\ limit on the X-ray luminosity derived by \citet{park05} is already lower than what can be expected from a newly born NS considering its thermal emission alone.

The aim of this work is to appraise the situation using currently available X-ray data. We derive new upper limits on the emission from a compact source in \src\ from recent \cxo\ observations, trying also to take into account the uncertainties due to the complicate environment and to the unknown pulsar's spectrum and rotational parameters. We then discuss the results in the context of the properties of neutron stars. In the following, unless the precise nature of the possible compact object left in \src\ is the focus of a sentence, we will use neutron star, pulsar and central source or object, more or less like synonyms.

\section{The \cxo\ observations}\label{new.analysis}
 
The \cxo\ \emph{X-ray Observatory} is the only instrument with good enough spatial resolution to resolve (partially) the structure of \src\ in X-rays. \cxo\ has two focal plane instruments: the micro-channel plate High Resolution Camera (HRC; \citealt{murray00}) and the Advanced CCD Imaging Spectrometer (ACIS; \citealt{garmire03}). The ACIS provides somewhat lower spatial resolution than the HRC and in imaging mode its readout speed is inadequate to sample the period of a fast-spinning pulsar, but it has a much larger effective area, especially at high energies, and is therefore better suited for our purpose. In fact, the effective area of the HRC drops rapidly by a factor $\sim$4 after the peak at 1\,keV and above 2\,keV is several time smaller than that of the ACIS. Since the opacity of the envelope to the high-energy emission is expected to decrease with time approximately as $\propto t^2$ (e.g. \citealt{mccray93,chevalier94,perna08}), here we used only the data from three of the longest and most recent \cxo\ observations in the archive (see Table\,\ref{obslog} and \citealt{frank16} for their details, and \citealt{park04,park05,ng09} for limits from older observations).

\begin{deluxetable}{lccc}
%\tabletypesize{font size command}
%\rotate
%\tablewidth{100pt}
%\tablenum{text}
%\tablecolumns{100pt}
\tablecaption{Log of the \cxo\ observations used in this work.\label{obslog}}
\tablehead{
\colhead{Instrument} & \colhead{Obs.ID} & \colhead{Date} & \colhead{Exposure} \\%& \colhead{Upper limit\tablenotemark{a}}\\
\colhead{} & & \colhead{dd-mm-yyyy} & \colhead{(ks)} %& \colhead{(counts s$^{-1}$)}
}
\startdata
ACIS-S/HETG& 15809 & 13-03-2014 & 70.5 \\%& $<$$2.0\times10^{-3}$\\
HRC-S/LETG & 16757 & 14-03-2015 & 67.7 \\%& --\\
ACIS-S/HETG & 16756 & 17-09-2015 & 66.6 \\%& $<$$2.6\times10^{-3}$ \\
\enddata
%\tablenotetext{a}{Upper limit on the 3--8~keV rate at 3$\sigma$ confidence level.}
\end{deluxetable}

The data were processed and analysed with the \cxo\ Interactive Analysis of Observations software package (CIAO version 4.8; \citealt{fruscione06}) and the calibration files in the CALDB database (version 4.7.1.). 
In all the observations, the instruments were operated with the grating spectrometers, the  Low Energy Transmission Grating (LETG) for the HRC and the High Energy Transmission Grating (HETG) for the ACIS. All the analysis presented here is based on the zero-order images, count rate and spectra. We note that the significant advantage of the ACIS over the HRC in terms of effective area for hard photons remains also when the transmission gratings are used.\footnote{See for example \url{http://cxc.harvard.edu/caldb/prop_plan/pimms/}.}
For the ACIS data, we removed the pixel randomization added by the \cxo\ software and used the energy-dependent sub-pixel event repositioning (EDSER) algorithm by \citet{li04} to achieve sub-pixel resolution. 
The HRC data were used mainly to check that the procedure did not produce image artefacts. 

\subsection{Analysis of the ACIS data}\label{spectra}

\begin{figure*}
\centering
\resizebox{\hsize}{!}{\includegraphics[angle=0]{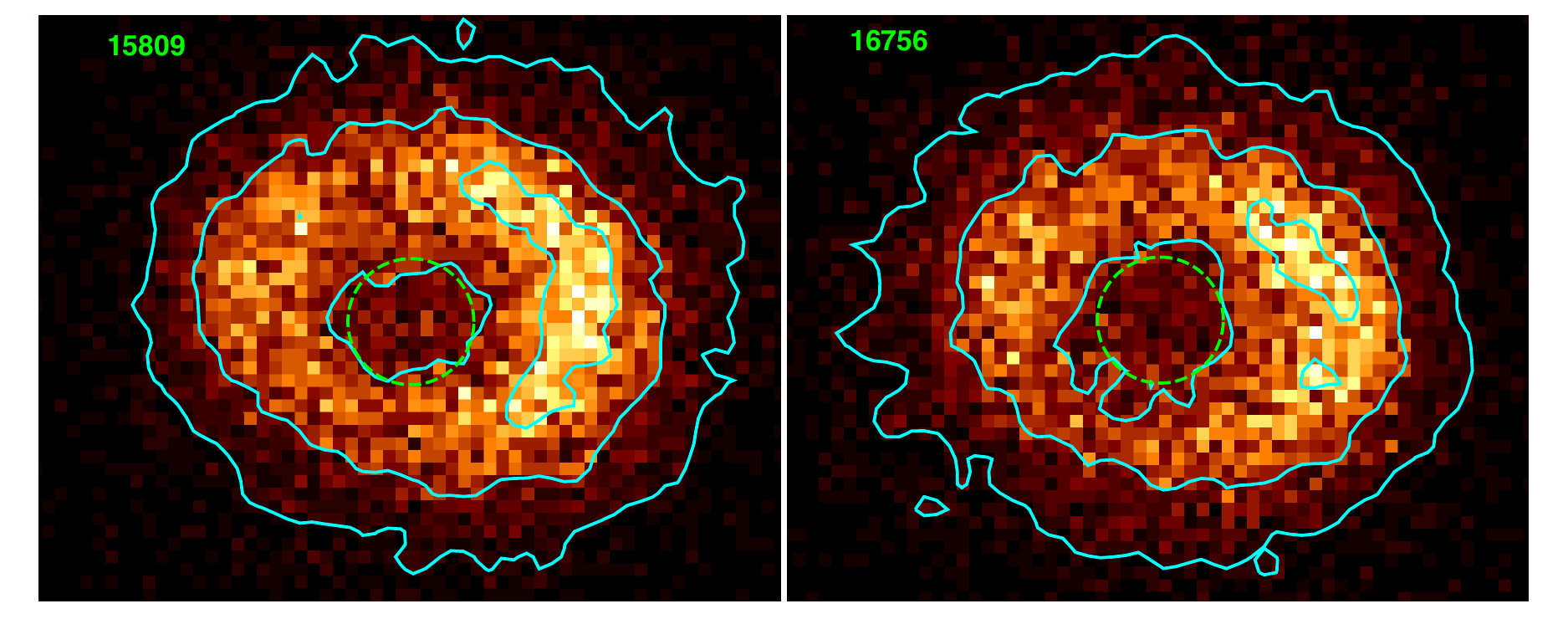}}
\caption{ACIS 2014 and 2015 images of \src\ in the 0.3--8~keV band with subpixel binning (1/8 of the native pixel size); the Obs.ID is indicated in each panel. The  dashed circles show the area we considered to evaluate the limits (approximately $0\farcs3$ radius). The brightest drawn contour levels in each panel correspond to $\sim$25 counts bin$^{-1}$ and the other levels are spaced by a factor of $\sqrt{2}$. \label{acis_images}}
\end{figure*}

In the X-ray band, at the time of the observations we considered (2014--2015, see Table\,\ref{obslog}), \src\ could be enclosed in a $\approx$3$''$ by 4$''$ (axes) ellipse (see Fig.\,\ref{acis_images}). The innermost structure, the `equatorial ring' (radius $R\approx0\farcs4$, equivalent to $\sim$0.1\,pc at 50\,kpc), has been interacting with the ejecta for many years \citep{frank16} and is rather bright in X-rays (Fig.\,\ref{acis_images}).

The main aims of the analysis described in this section are to see whether there are reasons to suspect that part of the X-rays originate from a pulsar and to measure the absorption in its direction. For the two ACIS observations we extracted the spectra from an inner circular region with radius of $0\farcs3$ (Fig.\,\ref{acis_images}). 
This choice, which entails the use of only a small fraction of the photon collected with \cxo, is motivated by the fact that for reasonable assumptions on the speed of a neutron star (projected velocity $<$2,000\,km\,s$^{-1}$, see e.g. \citealt{hobbs05}), the compact source must be within this radius. The background spectra were extracted from an annulus with radii of $2''$ and 4$''$, well outside the X-ray-bright rim of the supernova. This selection resulted in 0.5--8\,keV spectra of 414 photons for the first observation [net source count rate of $(5.8\pm0.3)\times10^{-3}$~counts\,s$^{-1}$, for a total of $411\pm20$ source counts and a signal-to-noise ratio of 99.4\%] and 325 photons for the second observation [net source count rate of $(4.8\pm0.3)\times10^{-3}$~counts\,s$^{-1}$, $323\pm18$ net counts and signal-to-noise of 99.3\%].

The spectra can be described by a model with one or more shock components (we used pshock in XSPEC / xspshock in Sherpa) modified for the absorption. Similarly to previous analyses (e.g. \citealt{zhekov06}), we found that the fit with a one-shock model results in an unreasonably low value of the absorption column density ($\sim$$9\times10^{19}$~cm$^{-2}$), consistent with zero and much lower that both the average total Galactic column density towards the LMC ($1.6\times10^{21}$~cm$^{-2}$; \citealt{kalberla05}) and the density measured in the direction of \src\ [$\sim$$(2$--$3)\times10^{21}$~cm$^{-2}$; \citealt{fitzpatrick90,michael02,park04,kalberla05}]. Additionally, fixing the \nh\ to a more plausible value of $2\times10^{21}$~cm$^{-2}$, we obtained a worse fit (although, still statistically acceptable), with a reduced $\chi^2$ ($\chi^2_\nu$) that increased from 1.18 for 29 degrees of freedom (dof) to 1.43 for 30 dof.
On the other hand, a two-shock model yields an absorption, as well as temperatures, in agreement with previous works. To obtain a better estimate of the magnitude of the absorption,\footnote{Considered the relatively large uncertainties and that the two observations were taken only 18 months apart, it seems reasonable to us to assume the same \nh\ value for the two spectra.} we fit simultaneously the two spectra with a numerical factor to account for the different fluxes (the results of the simultaneous fit are consistent with those of the individual fits). We obtained $N_{\rm H}=(2.6^{+0.8}_{-0.9})\times10^{21}$~cm$^{-2}$ (with the solar system abundances by \citealt{anders89} and the photoelectric absorption cross sections by \citealt{balucinska92}), $kT_1 = 0.7^{+0.4}_{-0.2}$~keV and $kT_2= 2.8^{+1.5}_{-0.5}$~keV, with $\chi^2_\nu= 0.88$ for 27 dof. The observed 0.5--10\,keV fluxes are $\sim$$1.8\times10^{-12}$ and $1.6\times10^{-12}$~\flux\ for Obs.IDs 15809 and 16756, respectively. The photon statistics of the spectra is too low to fit the elemental abundances in the region, but we verified that an acceptable fit ($\chi^2_\nu=1.02$ for 27 dof) and essentially identical results are found using the vpshock XSPEC model with the abundances fixed at the values of \citet[][table\,1]{zhekov06}, who measured shock temperatures of 0.5 and 2.7\,keV. We note that the use of a single absorption component with solar abundances is an oversimplification, since at least a Galactic and a LMC component should be considered; however, here we are not interested in an absolute measurement of the column density: our purpose is to parameterize the absorption with a simple indicator that will be used in the following sections. Finally, by using in Sherpa the jdpileup pileup model by \citet{davis01}, we estimated in both observations a pileup fraction lower than 1\%, so no attempt to correct for it was made in the following. 

\subsection{Upper limits on the X-ray emission of a central point source}

To set the most conservative upper limits on the luminosity of a central source, one should assume that all the flux observed inside the central ring is produced by the central source. However, the spectral analysis of Sect.\,\ref{spectra}, with the low absorption and the shock components, strongly suggests that most of the X-ray luminosity is produced by the shocked circumstellar matter and supernova debris and does not come from the innermost region of the remnant, where the pulsar would be expected to reside. In light of this, we chose to estimate the upper limit as the 3$\sigma$ noise level (evaluated from the Poisson fluctuations of the background) in the $0.3$-arcsec-radius region at the center of \src. This turned out to be $3.9\times10^{-4}$~counts\,s$^{-1}$ in observation 15809 and $4.1\times10^{-4}$~counts\,s$^{-1}$ in observation 16756 (2--8~keV), $\approx$30\% of the 2--8~keV counts in both cases. We checked that point sources simulated with ChaRT and MARX \citep{carter03,davis12} with those rates at the center of the supernova are detected in the X-ray images at the expected confidence level (in the simulations we adopted various thermal and nonthermal models, as well as different absorption levels, as discussed in Sections\,\ref{thermal} and \ref{nonthermal}, but the results were virtually independent of the specific spectral shape).\footnote{Strictly speaking, both methods yield an estimate of the sensitivity of the observation to the flux from a point source rather than an upper limit on it; however, considered the comparatively high number of photons and for the aims of this study, the two related quantities can be considered equivalent.} % (but we note that we are at the limit of the \cxo\ spatial resolution).
As a further test of consistency of the different pieces of information, we fit to the 0.5--8\,keV spectra a model consisting of a shock component with absorption fixed at  $N_{\rm H}=2.6\times10^{21}$~cm$^{-2}$ plus a power law with photon index fixed at $\Gamma=2.1$ (which is the model that describes the emission of the Crab pulsar in the soft X-ray band; \citealt{kirsch05}) modified by a second, independent absorption component (free to vary), so to see if the emission expected from a young pulsar is compatible with the available data. We obtained an acceptable fit and a reasonable shock temperature ($\chi^2_\nu=1.05$ for 28 dof and $kT=(1.5\pm0.2)$\,keV) with the `pulsar component' accounting for $\approx$25\% of the total emission ($<$65\% at 3$\sigma$) and a 3$\sigma$ upper limit on the local absorption of $N_{\rm H}<1.8\times10^{23}$~cm$^{-2}$.

To convert the ACIS count rate limits into luminosity limits (using XSPEC and the ancillary response files for the spectra extracted from the 0.3-arcsec-radius regions to correct for the PSF and effective area fractions), several hypotheses and assumptions are necessary. We discuss them in the following sections.

Even though the bulk of the counts collected by the HRC are below 2\,keV (because of the thermal spectrum and the effective area curve of the detector), for the sake of diligence, we also searched its events within 0.3\,arcsec (around 1,100 photons in the 0.1--10 keV\,band) for coherent pulsations between 0.5\,ms and 1\,s. No statistically significant periodic signal was found and the upper limits on the pulsed fraction are not constraining (they are larger than 100\%).

\subsubsection{Thermal emission}\label{thermal}

Let us start by considering the thermal emission that arises from the entire surface of the star due to initial cooling.
For a newly-born neutron star, most cooling curves indicate a surface temperature of $T_{\mathrm{eff}}\approx 2.7\times10^6$\,K (e.g. \citealt{yakovlev04,aguilera08,page09,vigano13}), which is equivalent to $kT_{\mathrm{eff}}\approx0.23$~keV. This value can be considered an upper limit, since the temperature would actually be lower if fast neutrino cooling processes were present \citep[e.g.][]{yakovlev04,page09}. However, for prudence and to assess better the situation, we explored also the possibility of higher temperatures, up to 0.5\,keV.
If for the radiating surface we take a neutron star radius of 12~km \citep{lattimer17}, the bolometric luminosity is between $L\simeq5.5\times10^{34}$ and $1.2\times10^{36}$\,\lum\ for $kT_{\mathrm{eff}}\approx0.23$ and 0.5\,keV, respectively.\footnote{A `color-correction' factor $f_{\mathrm{c}}=T_{\mathrm{BB}}/T_{\mathrm{eff}}$ is usually used to take into account the distortion due to the stellar atmosphere in the spectrum emitted by the neutron star surface and connect it to the observed blackbody with temperature $T_{\mathrm{BB}}$. For the X-ray flux, $F_{\mathrm{X}}\propto T_{\mathrm{BB}}^4 (R_{\mathrm{BB}}/D)^2 = T_{\mathrm{eff}}^4 (R_\infty/D)^2$, with $R_\infty = R_{\mathrm{BB}}f_{\mathrm{c}}^2 = D_{\mathrm{10\,kpc}}\sqrt{k}f_{\mathrm{c}}^2$, where $R_{\mathrm{BB}}$ is the observed blackbody radius, $D$ the distance to the source ($D_{\mathrm{10\,kpc}}$ when in units of 10\,kpc), and $k$ is the normalization of the XSPEC bbodyrad model.  Typical values are in the range $1\lesssim f_{\mathrm{c}}\lesssim1.8$ (e.g. \citealt{ozel13}). Since the color-correction factor acts so as to keep the luminosity constant correcting for the high temperatures observed, its presence would not affect our discussion.\label{notecc}} The corresponding 2--8\,keV unabsorbed fluxes are $\sim$$3.7\times10^{-15}$ and $3.4\times10^{-12}$~\flux, and the 2--8\,keV X-ray luminosity for the distance $D=53.7\pm3$~kpc \citep{mccray16} ranges from $L_{\mathrm{X}}\simeq1.3\times10^{33}$ to $4.7\times10^{35}$\,\lum.

\begin{figure*}
\centering
\resizebox{\hsize}{!}{\includegraphics[angle=0]{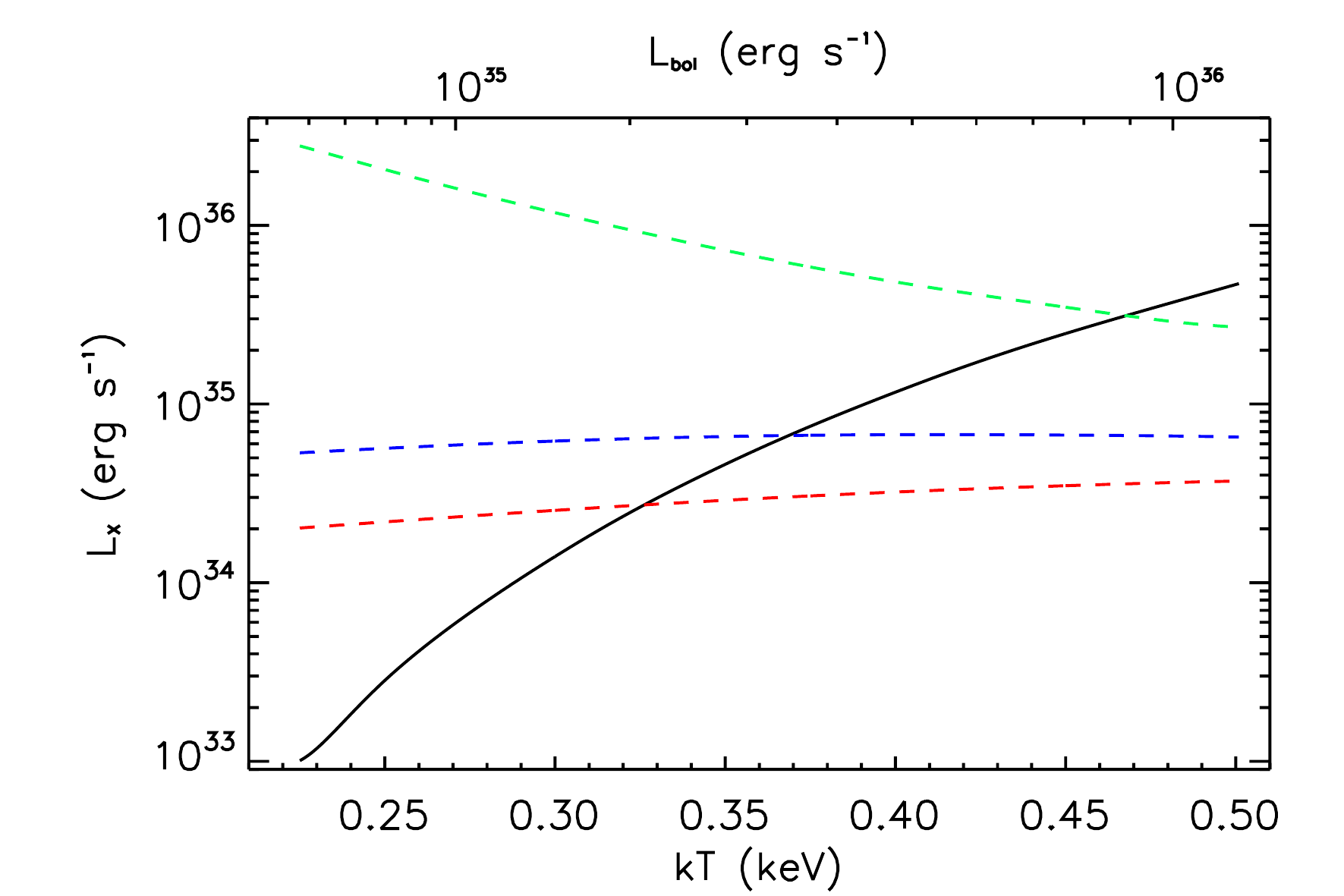}}
\caption{Constraints on the thermal emission from a neutron star. The solid black line shows the 2--8\,keV X-ray luminosity as a function of the blackbody temperature (see Section\,\ref{thermal}); the upper X axis show the corresponding bolometric luminosity. The dashed lines indicate the 3$\sigma$ upper limits obtained for different values of \nh: $3\times10^{21}$ (red), $3\times10^{22}$ (blue), and $1.8\times10^{23}$~cm$^{-2}$ (green).
\label{lbb}}
\end{figure*}

A crucial issue is the absorption of the X-rays coming from the centre of \src. The X-ray opacity is likely dominated by photoelectric absorption on inner-shell electrons of metals rather than by the Thomson scattering (see \citealt{corrales16}), which we shall neglect. Here we assume the envelope surrounding the compact object to be a sphere of uniform density  and homogeneous composition; for the moment, for simplicity we assume as metal abundance $Z=\mathrm{Z}_\odot$, so that we can use the values derived from the spectral analysis of the inner part of the remnant, but we shall discuss this later (Sect.\,\ref{alterednh}); we also neglect the possible ionization of the neutral matter by the central source. As the absolute minimum of the absorbing column, we posit the value derived from our X-ray fit, $2.6\times10^{21}$~cm$^{-2}$, which in the following we round off to $3\times10^{21}$~cm$^{-2}$; lacking more information, for the maximum we take the above-mentioned limit of $1.8\times10^{23}$~cm$^{-2}$. To have a rough order-of-magnitude reference value between these two extremes, we follow \citet{zanardo14}, who assume the presence of $M_{\mathrm{H}}\approx2.5\,\mathrm{M}_\odot$ of matter within a region $r\simeq0\farcs4$ (about 0.1\,pc), which is roughly the size of the equatorial ring (see also \citealt{blinnikov00,fransson13}). Correspondingly, the density is $4\times10^{-20}$\,g\,cm$^{-3}$ and \nh\ $\approx 3 M_{\mathrm{H}}/(4\pi m_{\mathrm{p}}r^2)\simeq3\times10^{22}$ cm$^{-2}$, where $m_{\mathrm{p}}$ is the proton mass.\footnote{In principle, one should add  to this value the Galactic absorption of $\sim$$(2$--$3)\times10^{21}$~cm$^{-2}$. However, given the considerable uncertainties in the estimate and the relatively small size of the change, the correction is unimportant.} The corresponding limits on the thermal emission for the different absorption levels are shown in Fig.\,\ref{lbb}. It is apparent that for the lower end of the range of temperature considered ($kT\lesssim0.3$\,keV) a purely thermally-emitting neutron star, even if present, would not have been detected even for the lowest conceivable \nh\ value.

\subsubsection{Nonthermal emission}\label{nonthermal}

Unless some mechanism suppressing the neutron star magnetospheric activity that converts a fraction of rotational energy into X-rays is at work, the high-energy emission of an isolated young pulsar is expected to have a large---likely dominating---contribution from a nonthermal component (e.g. \citealt{kaspi06}).
For the pulsars for which the main rotational parameters (period $P$ and slow-down rate $\dot{P}$) and the X-ray luminosity have been measured, there appears to exist a correlation between the latter quantity and the rotational energy loss $\dot{E}_{\mathrm{rot}}$ \citep{seward88,becker97}.
Albeit they have all a large scatter, several empirical $L_{\mathrm{X}}$--$\dot{E}_{\mathrm{rot}}$ relations have been derived from different surveys, samples of sources, etc (see \citealt{shibata16} and references therein); here we adopt that of \citet{possenti02}, $\log L_{\mathrm{X}} = 1.34\log \dot{E}_{\mathrm{rot}}-15.34$, which is valid over the 2--10\,keV range. Under the usual assumption of a magnetic dipole rotating in vacuum, 
$\dot{E}_{\mathrm{rot}}=B^2 \sin^2\theta\Omega^4R^6/(6c^3)\simeq3\times10^{43}B_{14}^2P_{10\,\mathrm{ms}}^{-4}$\,\lum, where $\Omega=2\pi/P$, $P_{10\,\mathrm{ms}}$ is in units of 10\,ms, $R=12$\,km, and for the angle between the magnetic and spin axes we take $\theta=\pi/2$. For the spectral model, the natural choice is to use a power law with photon index fixed at $\Gamma=2.1$, the value measured for the pulsar in the Crab nebula (PSR\,J0534+2200; $L_{\mathrm{X}}=1.1\times10^{37}$\,\lum), which is the best studied young pulsar we currently know of (but note that essentially all the young rotationally-powered pulsars known have similar spectral shape and slope; e.g. \citealt{becker97,possenti02,gotthelf03}).
\begin{figure*}
\centering
\resizebox{\hsize}{!}{\includegraphics[angle=0]{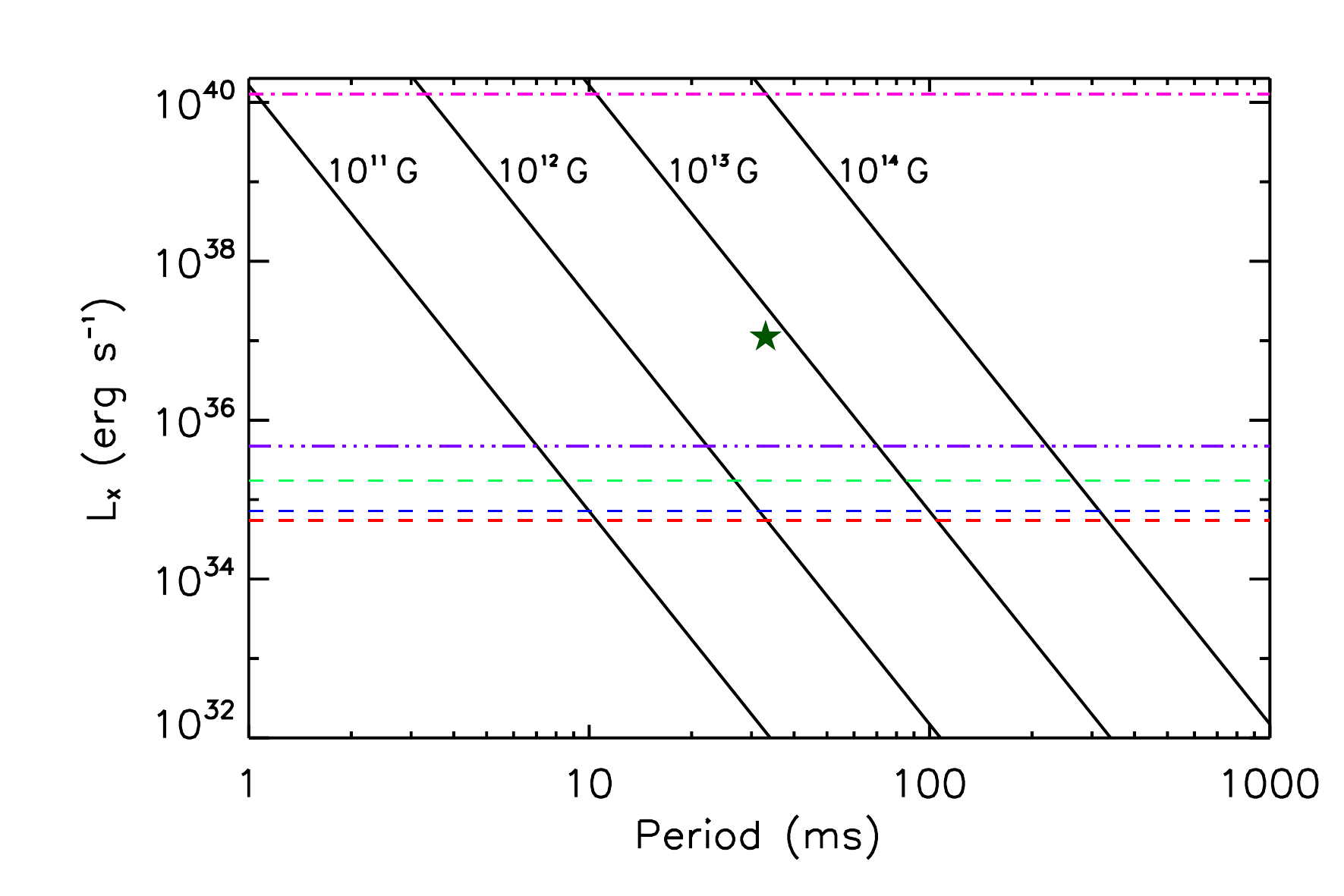}}
\caption{Constraints on the nonthermal emission from a neutron star. The solid black lines show the 2--8\,keV X-ray luminosity as a function of the period for a few values of the magnetic field, as indicated by the labels. The dark green star marks the position of the Crab pulsar. The dashed lines indicate the 3$\sigma$ upper limits obtained for different values of \nh: $3\times10^{21}$ (red), $3\times10^{22}$ (blue), and $1.8\times10^{23}$~cm$^{-2}$ (green). The three-dot-dashed line (violet) shows the extrapolation of the 2$\sigma$ upper limit derived with \igr\ in the 20--60 keV band. The purple dot-dashed line indicates the limit derived for the absorption with abundances similar to those of the ejecta (see Section\,\ref{alterednh}).
\label{lmax}}
\end{figure*}

In Fig.\,\ref{lmax} we show the limits for the same \nh\ values discussed for the thermal emission, together with the nonthermal X-ray luminosity expected for some combinations of the neutron star period and magnetic field. While a pulsar akin that in the Crab nebula is not compatible with the data,  there is ample room for viable combinations of parameters (e.g. $B=10^{12}$\,G and any period $P>25$\,ms).

\subsubsection{The shielding curtain}\label{alterednh}

Now we devote some attention to the composition of the matter in the curtain screening the site of the possible compact object, where metals that have a  large cross section for photoelectric absorption for photons with energy of $\gtrsim$1 keV,  such as C, O, Si, and Fe \citep{morrison83}, can be expected to be overabundant. We used the ejecta composition in table\,1 of \citet{dessart10}, which is based on hydrodynamical models by \citet[and references therein]{woosley02}, to create an abundance table for the photoelectric absorption model in XSPEC (Table\,\ref{abund}). For the elements for which they do not provide information, we fixed the abundance at the solar value by \citet{anders89}. 
We stress that this is not an attempt to model properly the absorption, but only to get an idea of how a non-standard composition can impact the limits.

Using the chemical mix in Table\,\ref{abund}, the \nh\ of $3\times10^{22}$~cm$^{-2}$ that we derived from the assumption of a local density of $4\times10^{-20}$\,g\,cm$^{-3}$ becomes $\approx$$7.6\times10^{21}$~cm$^{-2}$. Note that the lower nominal \nh\ value only reflects the smaller H fraction in the mix and, with the abundances in Table\,\ref{abund}, it actually results in a much larger X-ray abatement. In fact, with this absorption (again, we neglected the Galactic absorption component), the upper limit on the nonthermal luminosity of a pulsar derived from the data is $\approx$$1.3\times10^{40}$\,\lum\  (Fig.\,\ref{lmax}), which is much larger than the luminosity of the Crab pulsar. The upper limit on the thermal component is even less binding ($L_{\mathrm{X}}>10^{41}$\,\lum).

\begin{deluxetable}{lccc}
\tablecaption{Abundances adopted. The solar system values are from \citet{anders89} and all abundances are relative to H.\label{abund}}
\tablehead{
\colhead{Element} & \colhead{Solar system $Z/H$} & \colhead{Custom $Z/H$}\\ 
%& \colhead{\footnotesize\color{red}column to be removed} & 
}
\startdata
H    & 1.00 &1.00	       \\
He   & 9.77e--02 &2.29	       \\
Li   & 1.45e--11 &1.45e--11	       \\
Be   & 1.41e--11 &1.41e--11	       \\
B    & 3.98e--10 &3.98e--10	       \\
C    & 3.63e--04 &1.55e--01	       \\
N    & 1.12e--04 & 6.85e--03	       \\
O    & 8.51e--04 &3.05e--01	       \\
F    & 3.63e--08 &3.63e--08	       \\
Ne   & 1.23e--04 & 8.63e--02	       \\
Na   & 2.14e--06 & 2.14e--06	       \\
Mg   & 3.80e--05 &1.52e--02	       \\
Al   & 2.95e--06 &2.95e--06	       \\
Si   & 3.55e--05 & 2.03e--02	       \\
P    & 2.82e--07 &2.82e--07	       \\
S    & 1.62e--05 & 7.61e--03	       \\
Cl   & 3.16e--07 &3.16e--07	       \\
Ar   & 3.63e--06 & 1.27e--03	       \\
K    & 1.32e--07 &1.32e--07	       \\
Ca   & 2.29e--06 & 1.02e--03	       \\
Sc   & 1.26e--09 &1.26e--09	       \\
Ti   & 9.77e--08 &1.65e--05\\
V    & 1.00e--08 &1.00e--08	       \\
Cr   & 4.68e--07 &4.68e--07	       \\
Mn   & 2.45e--07 &2.45e--07	       \\
Fe   & 4.68e--05 &2.54e--03	       \\
Co   & 8.32e--08 &8.32e--08	       \\
Ni   & 1.78e--06 & 2.13e--02	       \\
Cu   & 1.62e--08 &1.62e--08	       \\
Zn   & 3.98e--08 &3.98e--08	       \\
\enddata
%\tablenotetext{a}{The solar system abundances are from \citet{anders89}.}
\end{deluxetable}

\section{Discussion and conclusions}

The mass of Sk\,$-$69$^\circ$\,202 and the burst of neutrinos that accompanied the explosion suggest that a neutron star was formed in \src\ by the process of core collapse (although it cannot be excluded that it further collapsed in a black hole if enough fallback material piled upon its surface; e.g. \citealt{zampieri98}). However, thirty years after the explosion and despite observations in every band of the electromagnetic spectrum, there is still no positive evidence for a compact object of any kind in the remnant. 

Optical and ultraviolet (UV) searches were performed for both periodic signals and point-like emission. After some claims of detection of pulsations that were later retracted or not confirmed by subsequent observations or reanalysis, upper limits on the pulsed emission for periods between 0.2 and 10~s were set with a limiting $V$ magnitude of $\sim$24.6 using \hst\ and the ground-based Anglo-Australian Telescope \citep{percival95,manchester96}. 
%\citet{graves05} derived with \hst\ UV/optical limits of a few $10^{33}$--$10^{34}$~\lum\ assuming a dust absorption in the remnant $\leq$97\%. However, the site of the pulsar is probably cloaked in a cloud of impenetrable gas and recent observations in the far-infrared (IR) and in the sub-mm continuum with \emph{Herschel} and ALMA \citep{matsuura11,matsuura15,matsuura17} have shown that there is a lot of dust in the ejecta, probably enough to make them completely opaque in the near-IR and optical bands.
\citet{graves05} assumed an attenuation due to the dust absorption in the remnant $\leq$97\% and derived with \hst\ UV/optical limits on the luminosity of a compact remnant of a $\rm few \times 10^{33}$ to $10^{34}$~\lum. Recent observations in the far-infrared (IR) and in the sub-mm continuum with \emph{Herschel} and ALMA \citep{matsuura11,matsuura15,indebetouw14} showed the presence of substantial amount of dust in the ejecta. If the dust is distributed in clumps, some light could scatter around the clumps and the extinction could be lower than the limit assumed by \citet{graves05}, at least in some lines of sight. However, if nearly half a solar mass of dust \citep{matsuura11,matsuura15,indebetouw14} fills the ejecta uniformly, even higher extinction might be possible and the site of the pulsar would be cloaked in a cloud impenetrable to the IR/optical light.

A limit on the luminosity of the alleged central source can also be derived by comparing the bolometric emission of the remnant with the power injected by titanium-44 ($^{44}$Ti) decay (\citealt{grebenev12,boggs15}; \citealt{mccray16}). The decay of the $^{44}$Ti  at 10,000 days is expected to deposit energy at a rate of $L_{\rm ^{44}Ti}\sim280$\,L$_\odot$, most of which is radiated at IR wavelengths from a population of $\sim$0.5\,M$_\odot$ of dust grains \citep{matsuura11,indebetouw14} with luminosity $L_{\mathrm{dust}}\sim220$\,L$_\odot$. While there seem to be not much room for an additional energy input from a central source, one must consider that (i) both the $^{44}$Ti energy deposition and the IR luminosity have uncertainties of the order of 20 per cent or more and (ii) most of the X-ray flux from the central source would be absorbed via photoionization, the majority of the energy ($\approx$75\%) being used to photoionize and only the remaining ($\approx$25\%) available for heating of the dust particles.\footnote{We considered a power-law photon spectrum with index $\alpha$, $\frac{d n}{d\nu}=N_0(\frac{\nu}{\nu_0})^{-\alpha}$, and we approximated the photoionization cross section as $\sigma(\nu)=\sigma_0(\frac{\nu}{\nu_0})^{-3}$. The ratio of the energy in heat over the total photoionization energy is $f=\frac{E_{\mathrm{thermal}}}{E_{\mathrm{total}}}=\frac{\int_{\nu_0}^{+\infty}(h\nu-h\nu_0)\sigma(\nu)\frac{d n}{d\nu}d\nu}{\int_{\nu_0}^{+\infty}h\nu\sigma(\nu)\frac{d n}{d\nu}d\nu}=\frac{1}{\alpha+2}$. For $\alpha=0$, half of the energy goes into heating. For a more reasonable $\alpha=2$, only 25\% of the absorbed radiation is turned into heat.} We conclude that the comparison of the dust emission with the $^{44}$Ti energy deposition rates is consistent with the presence of a central source with a luminosity of $L_{\mathrm{dust}}-fL_{\rm ^{44}Ti}\gtrsim100$\,L$_\odot$ ($\sim$$4\times10^{35}$\,\lum).

Repeated observations in radio at different frequencies provided limits on the flux density $<$115\,$\mu$Jy for pulsed emission \citep{manchester07}. These limits are not particularly constraining however, because of the large distance to the LMC (e.g. \citealt{manchester05}). Furthermore, the non detection could be due to free--free absorption in the supernova remnant \citep[see also][]{wang17}, or simply to an unfavourable beaming. Indeed, there are numerous young and energetic pulsars (including and, seemingly, mostly rotation-powered ones) that are not detected as radio pulsars (e.g. \citealt{caraveo14}).

In X-rays, the deepest upper limits on the emission of a point source can be obtained using \cxo. We believe that, when the substantial uncertainties involved in the X-ray analysis (in particular, in the absorption) are considered, the limits are not particularly restrictive even in this band. The thermal component of the emission of a pulsar would easily escape detection in the available data sets; in particular, for the lower temperatures in the range of what can be expected from a `baby' neutron star ($kT\lesssim0.3$\,keV), the limits are not constraining even in the case of the lowest conceivable absorption, corresponding to the total Galactic \nh\ (with solar-system abundances) along the line of sight towards \src\ (Fig.\,\ref{lbb}). 

In the case of the nonthermal emission, the situation is more critically dependent on the absorption. A pulsar as bright in X-rays as the one in the Crab nebula (and with a similar emission spectrum), should be detectable for the range of \nh\ that we explored if the composition of the absorbing matter is similar to that of the solar system. However, our exercise of altering the chemical composition of the absorber, so to reflect an enrichment of the elements that should be abundant in the ejecta of a massive star at the end of its life, shows that the X-ray limits become totally loose in the instance of very high metallicity. In that case, the Crab pulsar itself could be lurking in the remnant (Section\,\ref{alterednh}). Lacking sound information on the quantity and the composition (and ionization state) of the absorbing gas, the most reliable limits are probably those obtained in soft $\gamma$-rays. Using the IBIS/ISGRI hard-X-ray telescope on board  \igr, \citet{grebenev12} derived a 2$\sigma$ upper limit of $3\times10^{35}$\,\lum\ in the 20--60\,keV band for a power-law continuum with photon index $\Gamma=2.1$. This value, extrapolated to the 2--8\,keV band by assuming the Crab pulsar's spectrum, corresponds to $4.7 \times10^{35}$\,\lum, as shown in Fig.\,\ref{lmax}. This luminosity is higher than, but comparable to, the limits that we obtained assuming solar-system-like abundances: a $\rm few \times10^{35}$\,\lum\ (Fig.\,\ref{lmax}). It is also consistent with the limit derived from the reprocessed radiation from the $^{44}$Ti decay.
We note that for a pulsar, a nonthermal X-ray luminosity of \mbox{$\approx$$(1$--$5)\times10^{35}$\,\lum} corresponds to a rotational energy loss of $\approx$$(0.5$--$1.5)\times10^{38}$\,\lum.

Overall, it seems that while a Crab-like pulsar is incompatible with the observations (essentially, the $\gamma$-ray observations), there is ample room for the presence of an X-ray-emitting neutron star. In fact, many combinations of period and magnetic field plausible for an ordinary young neutron star are allowed by the limits in Fig.\,\ref{lmax}.
A recent work by \citet{gullon14}, for instance, found that the observed Galactic population of neutron stars is well reproduced by distributions of intial periods ($P_0$) in the range 0.1--0.5\,s, with only a small fraction of objects with $P_0<0.1$\,s, and initial magnetic field strength $\log B_0[\mathrm{G}]\approx13.0$--$13.2$ with width $\sigma(\log B_0) = 0.6$--$0.7$ (see also  \citealt{faucher06,popov10}).

While a supposed pulsar in \src\ does not necessarily have to be an `unusual' neutron star, an object similar to the so-called `central compact objects' (CCOs; e.g. \citealt{deluca17}) is certainly a very viable possibility. CCOs are steady X-ray sources with seemingly thermal spectra and no counterparts in radio and gamma wavebands; their periods, although measured in only a few sources, are in the 0.1--0.5 s range. These properties are clearly consistent with the observational constraints for \src, and CCOs seem to be relatively common in our Galaxy \citep{deluca08}.
The emerging scenario for CCOs is that of young neutron stars either born with a weak magnetic field ($B<10^{11}$ G) or with a normal field `buried' beneath the surface \citep{ho11,vigano12,gotthelf13}. In the latter hypothesis, the submergence of the magnetic field is the consequence of a stage of  hypercritical accretion of debris matter after the supernova explosion, a situation that could have taken place in \src\ \citep{vigano12}. Conversely, a magnetar, given their typically higher thermal luminosities and hotter thermal components with respect to normal neutron stars \citep{vigano13,perna13}, would be somewhat disfavoured.

Finally, a new interesting piece of information has been recently produced by \citet{zanardo14}, who reported on the possible detection with ALMA at \mbox{102--672 GHz} (after removing synchrotron emission modelled with ATCA at 44\,GHz) of a flat-spectrum region slightly westward of the SN site, whose properties are consistent with a pulsar wind nebula (PWN). They estimated $L_{\mathrm{PWN}}\approx5.4\times10^{33}$~\lum\ for frequencies between 102 and 672~GHz. If the PWN is powered by the pulsar with an efficiency of $\sim$1\%, the radio luminosity implies a rotational energy loss $\dot{E}_{\mathrm{rot}}\lesssim10^{35}$~\lum\ and hence an X-ray luminosity not larger than $L_{\mathrm{X}}\approx10^{32}$\,\lum\ \citep{zanardo14}. Even in the case the identification of the PWN is correct, we do not regard these limits as compelling, since the luminosity in the 102--672~GHz would be only a lower limit for its emission. Furthermore, the relationship between pulsar spin-down power and PWN luminosity is rather uncertain and, similarly to the $L_{\mathrm{X}}$--$\dot{E}_{\mathrm{rot}}$ relation for pulsars, has a large scattering (e.g. \citealt{mattana09}). However, this candidate PWN is the only hint of the presence of a pulsar in \src\ obtained so far. Further studies of this possible PWN or, more in general, the detection of a compact radio source with flat spectrum and/or polarized emission near the center of the supernova remnant, probably represent the best hope to establish the presence of a neutron star in \src\ in the next few years.

\acknowledgments
This research is based on observations made by the \cxo\ X-ray Observatory and has made use of software provided by the \cxo\ X-ray Center (CXC) in the application packages CIAO, ChIPS, and Sherpa. PE and NR acknowledge funding in the framework of the NWO Vidi award A.2320.0076. MM is supported by an STFC Ernest Rutherford fellowship (ST/L003597/1). RP was partly supported by NSF award AST-1616157. We thank the anonymous referee for many valuable comments. PE is grateful to Lia Corrales, Sandro Mereghetti,  Elisa Costantini and Mathieu Renzo for useful discussions. RP and PE are grateful to Richard McCray for relevant information.

\facility{\cxo\ (ACIS, HRC)} 
 \software{ChaRT \citep{carter03}, CIAO (v4.8; \citealt{fruscione06}), FTOOLS (v6.21; \citealt{blackburn95}), MARX (v5.3; \citealt{davis12}), Sherpa \citep{freeman01}, XSPEC (v12.9.1; \citealt{arnaud96})}
 
\bibliographystyle{aasjournal}
\bibliography{biblio}

\end{document}